\documentclass[aps,prd,nofootinbib,twocolumn,superscriptaddress,floatfix,notitlepage]{revtex4-1}
\pdfoutput=1

\usepackage{graphicx}
\usepackage{amsmath,amsfonts,amssymb}
\usepackage{color}
\usepackage[breaklinks,colorlinks,urlcolor=blue,citecolor=blue,linkcolor=magenta]{hyperref}
\usepackage{verbatim}
\usepackage{enumitem}
\usepackage{hyperref}

\newcommand{\td}{{\rm d}}
\newcommand{\vect}[1]{\boldsymbol{#1}}
\DeclareMathOperator{\artanh}{artanh}

\newcommand{\be}{\begin{equation}}
\newcommand{\ee}{\end{equation}}
\newcommand{\bea}{\begin{equation} \begin{aligned}}
\newcommand{\eea}{\end{aligned} \end{equation}}

\def\lsim{\mathrel{\raise.3ex\hbox{$<$\kern-.75em\lower1ex\hbox{$\sim$}}}}
\def\gsim{\mathrel{\raise.3ex\hbox{$>$\kern-.75em\lower1ex\hbox{$\sim$}}}}

\begin{document}

\title{The dark timbre of gravitational waves}

\author{Juan Urrutia}
\email{juan.urrutia@kbfi.ee}
\affiliation{Keemilise ja Bioloogilise F\"u\"usika Instituut, R\"avala puiestee 10, 10143 Tallinn, Estonia}
\affiliation{Department of Cybernetics, Tallinn University of Technology, Akadeemia tee 21, 12618 Tallinn, Estonia}

\author{Ville Vaskonen}
\email{ville.vaskonen@pd.infn.it}
\affiliation{Keemilise ja Bioloogilise F\"u\"usika Instituut, R\"avala puiestee 10, 10143 Tallinn, Estonia}
\affiliation{Dipartimento di Fisica e Astronomia, Universit\`a degli Studi di Padova, Via Marzolo 8, 35131 Padova, Italy}
\affiliation{Istituto Nazionale di Fisica Nucleare, Sezione di Padova, Via Marzolo 8, 35131 Padova, Italy}

\begin{abstract}
Gravitational wave timbre, the relative amplitude and phase of the different frequency harmonics, can change due to interactions with low-mass halos. We focus on binaries in the LISA range and find that the integrated lens effect of cold dark matter structures can be used to probe the existence of $M_{\rm v}\lesssim 10\, M_{\odot}$ halos if a single binary with eccentricity $e=0.3-0.6$ is detected with a signal-to-noise ratio $100 - 10^3$ and it is at $z_s=0.5$.
\end{abstract}

\maketitle

\section{Introduction} 

Gravitational waves (GWs) are often associated with sound because they are characterized by waveforms, and in general there is a poorer angular resolution than with electromagnetic signals. Like with sound, for GWs there is also the notion of timbre since eccentric binaries emit GWs in different harmonics simultaneously. The relative powers of the harmonics are determined by the binary eccentricity and the harmonic numbers~\cite{Peters:1963ux}.

GWs are sensitive to wave optics interactions with small halos, which can leave detectable imprints on the waveform~\cite{Takahashi:2003ix,Fairbairn:2022xln, Caliskan:2022hbu,Caliskan:2023zqm,Savastano:2023spl,Tambalo:2022wlm,Ma:2023jki,Cremonese:2021puh,Urrutia:2023mtk,Brando:2024inp} or a phase difference with respect to an electromagnetic counterpart~\cite{Takahashi:2016jom,Morita:2019sau,Ezquiaga:2020spg, Suyama:2020lbf}. Both methods rely on frequency-dependent effects for detectability. Moreover, in the case of strong lensing, the lensed image can deviate from unlensed waveforms in presence of higher harmonic modes, precession or eccentricity~\cite{Dai:2017huk,Ezquiaga:2020gdt}.

In this work, we propose a novel probe of the wave optics effects: measurements of the timbre of the GW signal from an eccentric binary. We study if the measurements of the timbre can be used to probe the low-mass end of the dark matter (DM) halo mass function (HMF) where deviations from the cold dark matter (CDM) predictions may appear. For example, the small-scale structures are suppressed in warm or ultralight DM models~\cite{Bode:2000gq,Hui:2016ltb,Rogers:2020ltq}. 

We focus on signals whose frequency does not change significantly. In the LISA sensitivity range, such signals can originate e.g. from white dwarf (WD) binaries, intermediate-mass black hole (IMBH) binaries or extreme mass ratio binaries~\cite{Barack:2006pq,Babak:2017tow,Berry:2019wgg,Gair:2017ynp,Hannuksela:2018izj,Bonetti:2020jku}. The signals from WD binaries are weak and only sources in the Milky Way can be seen individually. Such signals can get lensed by Milky Way substructures. As we are interested in lensing by DM halos, the most relevant sources are IMBH binaries that produce the strongest extragalactic signals.

We compute the integrated effect of the DM halo population and show that very light halos, which are integrated out in weak lensing studies as a constant density field~\cite{Bartelmann:1999yn}, induce changes in the amplitude and phases of the different harmonics of a mHz GW signal, effectively changing the timbre. The effect is detectable with LISA in the first harmonics, for example, if a binary is at $z_s=0.5$ with a signal-to-noise between $100-10^3$ and has eccentricity $e = 0.3 - 0.6$. We find that the effect mainly comes from halos of $\mathcal{O}(10\, M_{\odot})$. 

The reason for the dark timbre effect is that lensing by DM halos in wave optics causes a small frequency dependent enhancement of the GW amplitude that accumulates as the GW encounters several halos over cosmological distances. While the probability of detecting lensing by a single heavier halo is too low~\cite{Brando:2024inp,Fairbairn:2022xln}, the integrated effect of small halos is always present if such halos exist. The CDM model predicts a large abundance of light halos that have so far eluded observations~\cite{Zavala:2019gpq}. Therefore, the detection of dark timbre would provide a probe of the low-mass tail of the HFM and severely constraint deviations from CDM at small scales.

\section{Lensing by a single halo}  

Consider a binary whose orbital frequency remains almost constant during the observation. We denote the angular diameter distance of the binary by $D_s= d_{\rm c}(z_s)/(1+z_s)$, where $z_s$ is the corresponding redshift and $d_{\rm c}$ is the comoving distance, and assume that the GW signal emitted by the binary interacts with a halo at angular diameter distance $D_l$. In the frequency domain, the lensed waveform is $\tilde\phi_L(f) = F(f)  \tilde\phi(f)$, where $\tilde\phi(f)$ denotes the unperturbed GW signal. The amplification factor $F(f)$ under the thin-lens, Born, and eikonal approximations, is given by~\cite{schneider2012gravitational,Braga:2024pik} (see appendix~\ref{appA})
\be \label{eq:F}
    F(f) = \frac{w(f)}{2i \pi} \int \td^2 \vect{x} \,e^{i w(f) T(\vect{x},\vect{y})} \,.
\ee
The integral accounts for all the paths that the GW can take through it. The prefactor ensures that $F=1$ in the absence of lenses. The dimensionless vectors $\vect{x}$ and $\vect{y}$ are defined as $\vect{x} \equiv \vect{\xi}/\xi_0$ and $\vect{y} \equiv D_l \vect{\eta}/(D_s \xi_0)$, where $\xi_0$ denotes a characteristic length scale of the lens, and $\vect{\eta}$ and $\vect{\chi}$ are dimensionful vectors that give, respectively, the position of the source in the source plane and the position at which the GW crosses the lens plane. The dimensionless frequency $w$ and the dimensionless time delay function $T$ are given by
\begin{gather}   
    w(f) \equiv \frac{(1+z_l) D_s}{D_l D_{ls}} \xi_0^2 2\pi f \,,\\
    T(\vect{x},\vect{y}) \equiv \frac12 |\vect{x}-\vect{y}|^2 - \psi(\vect{x}) - \phi(\vect{y}) \,,
\end{gather}
where $D_{ls} = D_s - (1+z_l)/(1+z_s)D_l$ and $\phi(\vect{y})$ is defined such that $\min_{\vect{x}} T(\vect{x},\vect{y}) = 0$.

\begin{figure}
    \centering
    \includegraphics[width=0.95\columnwidth]{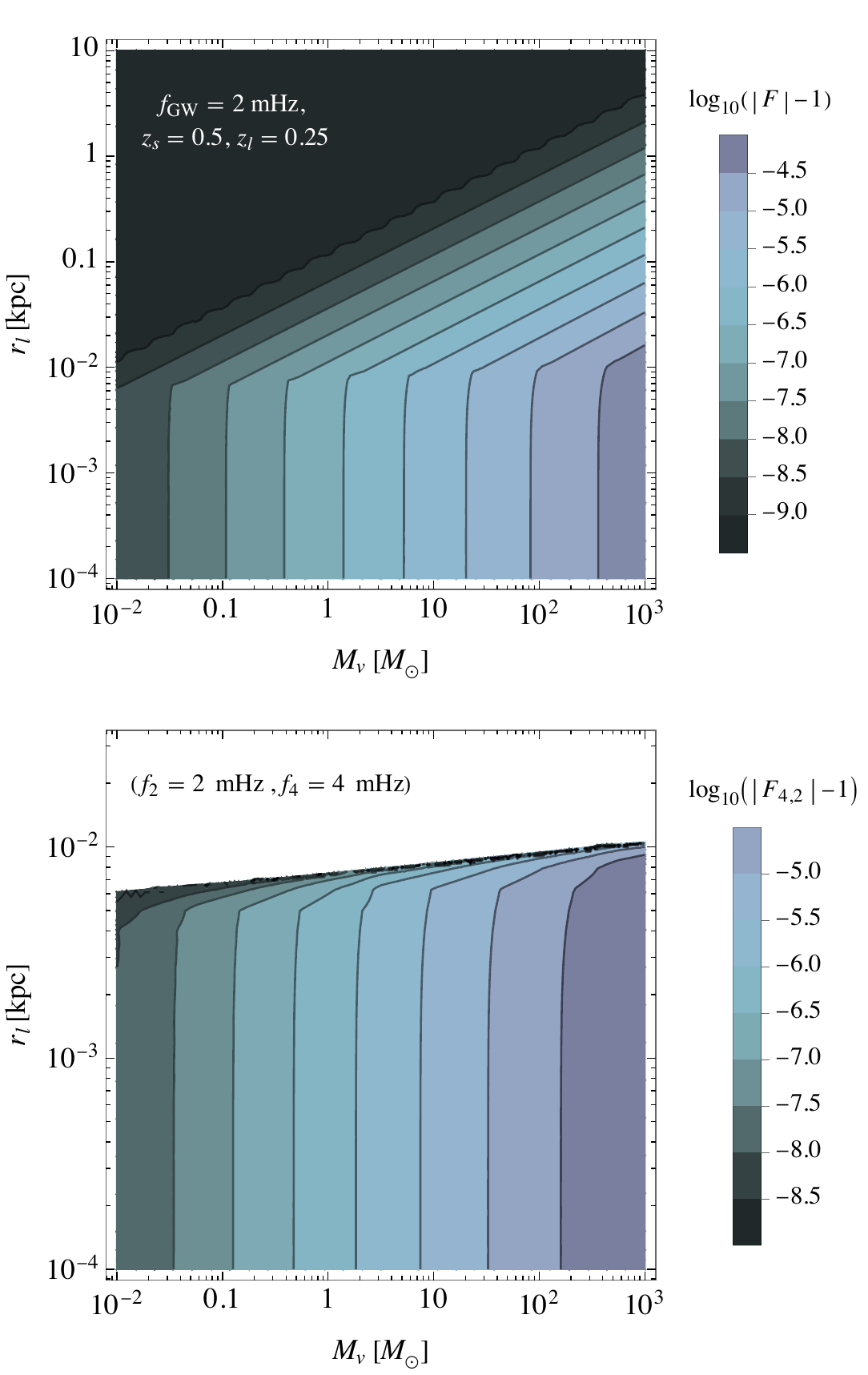}
    \caption{\textit{Upper panel:} The absolute value of the lensing amplification as a function of the halo mass and the projected distance of the source in the lens plane, for a source with $f_{\rm GW}=2\,{\rm mHz}$ at $z_s=0.5$ and a lens at $z_l=0.25$. \textit{Lower panel:} The magnitude of the relative amplification factor~\eqref{eq:Fij} for the same source-lens configuration as in the upper panel.}
    \label{fig:Amp}
\end{figure}

We approximate the halos by the Navarro–Frenk–White (NFW) profile $\rho(r) = r_s^3 \rho_s/[r(r_s+r)^2]$, which corresponds to the lens potential (see e.g.~\cite{Fairbairn:2022xln})
\be
    \psi(x) = \kappa \left[\ln ^2\!\left(\frac{x}{2 b}\right)-\artanh^2 \sqrt{1-\frac{x^2}{b^2}}\right] \,,
\ee
where $x=|\vect{x}|$ and $\kappa \equiv 2 \pi \rho_s r_s^3/M_{\rm v}$ and $b \equiv r_s/\xi_0$ are dimensionless parameters. The distances are scaled by $\xi_0 = R_E(M_v)$, where $R_E$ denotes the point mass Einstein radius, so the dimensionless frequency simplifies to $w = 8\pi (1+z_l) M_{\rm v} f$. The NFW profile is parameterized by the scale radius $r_s$ and the corresponding density $\rho_s$, which are both functions of the halo virial mass $M_{\rm v}$ through $\int_0^{R_{200}} \td r \, r^2 \rho(r) = 200 \rho_c R_{200}^3/3$ where $R_{200} = C(M_{\rm v}) r_s$ and $\rho_c$ denotes the critical density of the Universe. For the concentration parameter $C(M_{\rm v})$ we use the fit provided in~\cite{Ludlow:2016ifl}, which agrees with cosmological simulations in the relevant range of masses for this study~\cite{Wang:2019ftp}. For halos as light as the ones we will consider, interactions with stars can affect the profile. However, the survival rate is expected to be high~\cite{Delos:2019tsl,Hayashi:2002qv,vandenBosch:2017ynq,Taylor:2000zs,vandenBosch:2004zs,Berezinsky:2007qu,Penarrubia:2004et} so we can safely use the NFW profile. 

In the geometric limit, the GW follows a single classical trajectory, the minimal time path, around the NFW halo. The interaction lacks interference patterns and induces a frequency-independent amplification~\cite{Takahashi:2003ix},
\be \label{eq:amplification}
    |F| = \left(\det\left[ \partial_a \partial_b T(\vect{x}_j,y) \right] \right)^{-2}  \,,
\ee
where $y = |\vect{y}|$ and $\vect{x}_j$ corresponds to the classical path. However, in the wave optics regime, the amplification becomes frequency dependent, and numerical computation of Eq.~\eqref{eq:F} is necessary.\footnote{We do the numerical computation using the method introduced in~\cite{Takahashi:2004phd}.}

In the upper panel of Fig.~\ref{fig:Amp}, we show the amplification $|F|$ for different halo masses and distances in the lens plane. We have fixed the signal frequency to $f_{\rm GW} = 2\,$mHz, the source redshift to $z_s = 0.5$ and the lens redshift to $z_l = 0.25$. For a fixed halo mass in the wave optics regime, the interaction is independent of the projected position $r_l\equiv \xi_0 y$ of the source in the lens plane. In the geometric optics regime, the amplification effect rapidly decreases with $r_l$, roughly as $|F|-1 \propto r_l^{-4}$, but increases with the lens mass,\footnote{Linear scaling $|F|-1 \propto M_{\rm v}$ at a fixed frequency is expected if $\kappa$ and $b$ didn't change with $M_{\rm v}$. We find a slightly stronger scaling mainly because the compactness of the halos mildly decreases with decreasing $M_{\rm v}$.} roughly as $|F|-1 \propto M_{\rm v}^{7/6}$. The dividing distance between the wave optics and geometric optics regimes, on the other hand, increases very slowly with $M_{\rm v}$ remaining slightly below $r_l \approx 10$\,pc in the shown mass range.

Whereas a circular binary emits GWs only at twice its orbital frequency $f_{\rm orb}$, an eccentric binary emits at all integer multiples of it~\cite{Peters:1963ux}, 
\be
    f_n = n f_{\rm orb} \,.
\ee
These different harmonics get amplified by different amounts because the lensing effect is frequency-dependent. This is the dark timbre effect. In the lower panel of Fig.~\ref{fig:Amp}, we show the magnitude of the ratio of amplification factors that characterize the shift in the timbre, 
\be \label{eq:Fij}
    F_{i,j} \equiv \frac{|F(f_i)|}{|F(f_j)|}e^{i\Delta\phi_{i,j}} \,,  \quad \Delta\phi_{i,j} \equiv \phi_i - \phi_j \,,
\ee
between the second and the fourth harmonic of an eccentric binary with orbital frequency $f_{\rm orb} = 1$\,mHz. As expected, there is no difference in the geometric optics regime's amplification factor for high-impact parameters. Only the halos in the wave optics regime, $r_l < 10$\,pc, modify the timbre of the signal. The phase difference is similar in magnitude to the amplification factor ratio.

\section{Total lens effect} 

The effect of a single lens is characterized by the lens redshift $z_l$, the projected distance of the source in the lens plane $r_l$ and the halo mass $M_{\rm v}$. We estimate the modification of the timbre caused by all of the halos the signal encounters in the wave optics regime on its way from the source to the detector by generating configurations of the heaviest lenses and estimating the expected effect from the light halos.

\begin{figure*}
    \centering
    \includegraphics[width=0.85\textwidth]{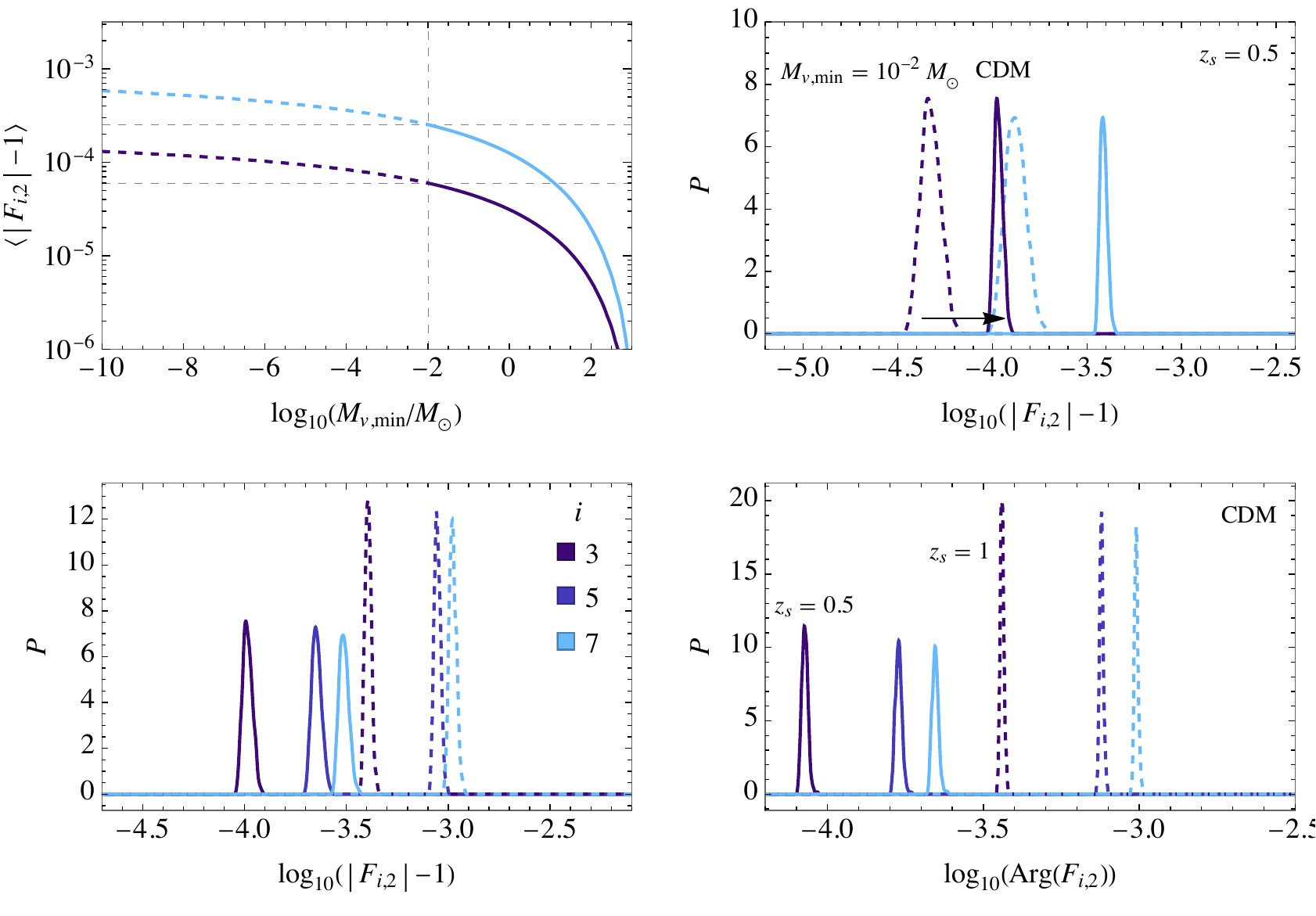}
    \caption{\textit{Upper panels:} The left plot shows the expected value for $|F_{i,2}|-1$ for the third (dark) and seventh (light) harmonic when integrating the halos from $M_{\rm v,min}$. The right plot shows how the distributions from the heavy halos, $M_{\rm v}>10^{-2}M_\odot$, (dashed curves) change when adding the contribution from the light ones. \textit{Lower panels:} The probability distributions of $|F_{i,2}|-1$ and ${\rm Arg}(F_{i,2})$ for $i=3,5,7$ and a source with $f_2 = 2\,$mHz at $z_s=0.5$ (solid) and $z_s=1$ (dashed).}
    \label{fig:plot2}
\end{figure*}

We consider only halos inside the wave optics regime, $r_l < r_{\rm max}(z_l, M_{\rm v})$, since those are the only ones causing a frequency-dependent effect. The total number of halos within that radius is
\be \label{eq:N}
    N = \int \!\td M_{\rm v} \int_0^{z_s} \!\!\td z_l \int_0^{ r_{\rm max}} \!\!\td r_l \frac{2\pi (1+z_l)^2 r_l}{H(z_l)} \frac{\td n(z_l)}{\td M_{\rm v}} ,
\ee
where $\td n(z_l)/\td M_{\rm v}$ is the HMF which we compute using the Sheth-Tormen approximation~\cite{Press:1973iz,Bond:1990iw,Sheth:1999mn} with the CDM power spectrum~\cite{Eisenstein:1997ik}. From Eq.~\eqref{eq:N}, we can identify the probability densities of the lens parameters:
\bea \label{eq:prob}
    &P\left(z_l,M_{\rm v}\right) = \frac{\pi (1+z_l)^2 r_{\rm max}^2}{N H(z_l)} \frac{\td n(z_l)}{\td M_{\rm v}} \,, \\
    &P\left(r_l|z_l,M_{\rm v}\right) = \theta\left(r_l-r_{\rm max}\right) \frac{2 r_l}{r^2_{\rm max}} \,.
\eea
We generate realizations of the lenses with mass $M_{\rm v} > 0.01 M_\odot$ that the signal encounters by sampling the probability distributions~\eqref{eq:prob} and approximate their total effect by taking a product of the amplifications caused by each of the lenses~(see appendix~\ref{appB}): 
\bea
    F_H(f) &\approx \prod_{j=1}^{N_{\rm h}} F(f|M_{{\rm v},j}, z_{l,j}, r_{l,j}) \\
    &\approx 1 + \sum_{j=1}^{N_h} \left[F(f|M_{{\rm v},j}, z_{l,j}, r_{l,j}) - 1\right] \,.
\eea
Here $N_{\rm h}$ is sampled from a Poisson distribution with mean $N$ in the mass range $M_{\rm v} > 0.01 M_\odot$ and the approximation holds because $F(f|M_{{\rm v},j}, z_{l,j}, r_{l,j})\approx 1$. We generate $10^3$ realizations to estimate the distribution of $F_H(f)$. 

For light halos with $M_{\rm v} < 0.01M_\odot$, the Poisson fluctuations are negligible because their number inside the wave optics regime is very large. Therefore, the effect they cause does not change significantly between different paths, and the distribution of the total lensing factor 
\be \label{eq:Ftot}
    F_{\rm tot}(f) = F_H(f) F_L(f) \approx F_H(f) + F_L(f) - 1 
\ee
can then be estimated by convoluting the distribution of $F_H$ with $\delta(F-F_L+1)$, where
\be \label{eq:FL}
    F_L(f) = \sum_j \left[1+\langle F(f)-1\rangle_{j}\, N_{{\rm h},j}\,(1-e^{-N_{{\rm h},j}})\right] 
\ee
is the expected effect of the light halos.\footnote{The HMF scales at small masses roughly as $\td n/\td\ln M_{\rm v} \propto M_{\rm v}^{-0.9}$. So, since $|F|-1 \propto M_{\rm v}^{7/6}$, the total amplification effect integrated over all halo masses is finite.} The sum in~\eqref{eq:FL} is over mass bins, $\langle F(f)-1\rangle_{j}$ is the expected deviation of the lensing factor from unity for halos in the mass bin $j$ and $N_{{\rm h},j}$ is the number of halos in the wave optics regime in that bin. This results in a translation of the distribution from heavy halos by adding the integrated effect of the light ones. 

\begin{figure}
    \centering    \includegraphics[width=0.94\columnwidth]{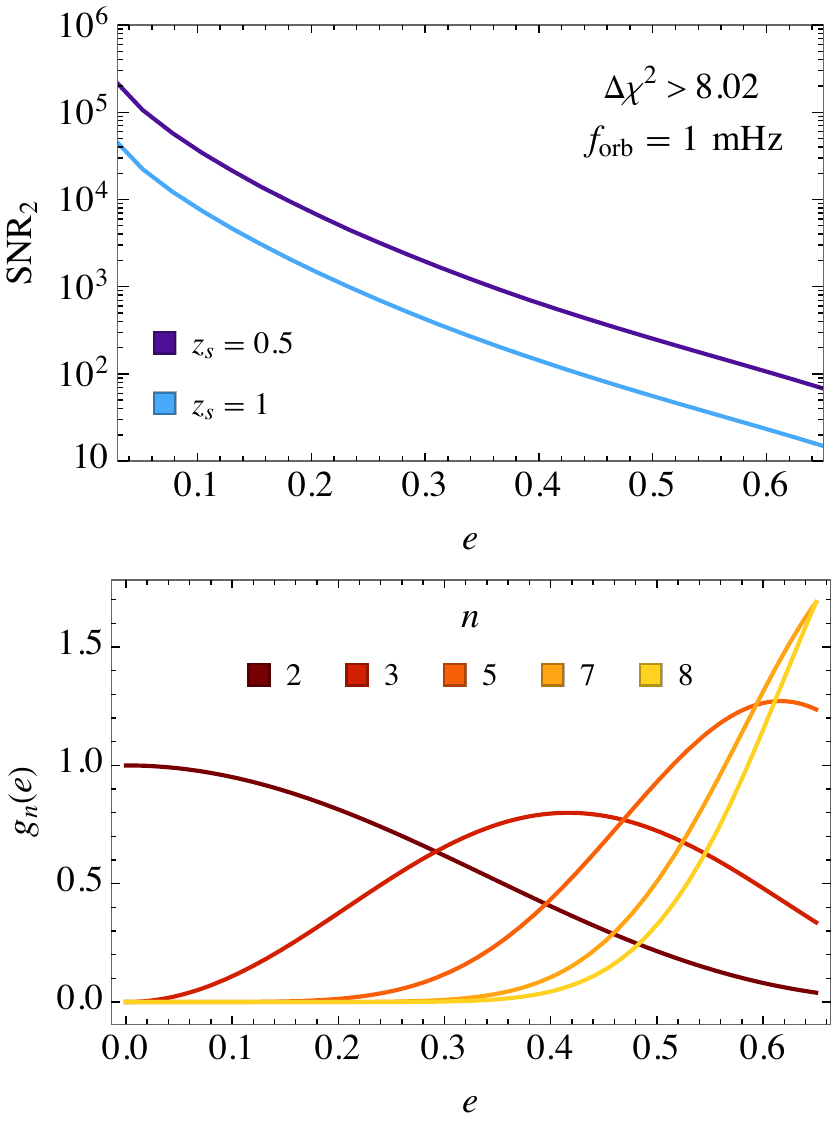}
    \caption{\textit{Upper panel:} The SNR of the second harmonic (${\rm SNR}_2$) necessary for detection of the shifted timbre as a function of the eccentricity of a binary orbiting at $1\,{\rm mHz}$. The dark line is for a source at $z_s=1$ and the light blue line for $z_s=0.5$. \textit{Lower panel:} The relative powers $g_n(e)$ of some of the lowest harmonics.}
    \label{fig:plot3}
\end{figure}

In the upper panels of Fig.~\ref{fig:plot2} we show the effect of adding the lightest halos. In the left panel, we can see that halos heavier than $100\, M_{\odot}$ do not contribute significantly. The reason is that they are too rare to fill the wave optics lensing tube. The main effect comes from halos in the range $0.01\, M_{\odot} -10\, M_{\odot}$. Below $0.01\, M_{\odot}$ the effect increases very slowly and we have explicitly checked that it converges. Halos lighter than $0.01\, M_{\odot}$ change the total lensing effect by an $\mathcal{O}(1)$ factor. In the right panel, we can see that the distributions are shifted by adding the contribution from the lightest halos.

In the lower panels of Fig.~\ref{fig:plot2} we show the probability distribution of the modulus and the argument of the amplification ratio between the second and the third, the fifth and the seventh harmonic for the same benchmark case as in Fig.~\ref{fig:Amp}, for $z_s=0.5$  (solid lines) and $z_s = 1$ (dashed lines). In the left panel, we see that the differences are of order $3\times 10^{-3}$ for $z_s=0.5$ and up to $10^{-3}$ for $z_s=1$. The distributions of the phase differences, shown in the lower right panel, scale similar to the distributions of the amplifications.

Previous GW microlensing studies have focused on searches of halos with $M_{\rm v} \gtrsim 10^6\, M_{\odot}$, which cause larger effects. However, for realistic DM halo profiles, such as Einasto or NFW, the effect in the geometric limit is simply a constant amplification~\cite{Fairbairn:2022xln} that is degenerate with the source properties. The lensing tube where wave optics effects are relevant is fixed by the frequency of the GW and the mass of the halo. For LISA frequencies and halos with $M_{\rm v}\gtrsim 10^6\, M_{\odot}$ the expected number of halos in the wave optics tube is very small and, consequently, it is not expected that LISA would see any such lensed events~\cite{Fairbairn:2022xln}. In this study, we instead consider much lighter halos $M_{\rm v} \lesssim 10\, M_{\odot}$ which are expected, assuming the CDM model, to induce wave optics effects in all of the GW events that LISA will see. The next section discusses the precision necessary to test and detect this effect.

\section{Detectability and prospects} 

To detect the dark timbre effect, we require that the signal with the shifted timbre cannot be fitted well with a template using the standard timbre. In general, the wellness of a template fit is given by~\cite{Jaranowski:2005hz} 
\be \label{eq:chi}
    \Delta\chi^2 = 4\min_{\vec\theta} \int \td f\frac{|\tilde{h}(f)-\tilde{h}_{\rm T}(f,\vec\theta)|^2}{S(f)} \,, 
\ee
where $\tilde{h}(f)$ refers to the GW signal and $\tilde{h}_{\rm T}(f,\vec\theta)$ the template parameterized by $\vec\theta$. For a binary that evolves very slowly, the waveform is approximately a sum of delta functions and Eq.~\eqref{eq:chi} simplifies to 
\bea
    &\Delta\chi^2 = 4 \min_{\lambda,\phi,e'} \sum_{n=1}^\infty \frac{A^2 |F(f_n) \sqrt{g_n(e)}- \lambda e^{i\phi} \sqrt{g_n(e')}|^2}{S(f_n)} \\
    &= \frac{{\rm SNR}_2^2}{g_2(e)} \min_{\lambda,\phi,e'} \sum_{n=1}^\infty \frac{S(f_2)}{S(f_n)} \left|F_{n,2} \sqrt{g_n(e)}- \lambda e^{i\phi} \sqrt{g_n(e')}\right| ^2 ,
\eea
where $A$ characterizes the amplitude of the signal and ${\rm SNR}_2$ is the signal-to-noise ratio of the second harmonic. The template is parameterized by an amplitude $\lambda \approx 1$, an eccentricity $e'\approx e$, and a phase $\phi \approx 0$. The function $g_n(e)$ gives the relative power radiated in the $n$th harmonic by a binary with eccentricity $e$~\cite{Peters:1963ux}. We show $g_n(e)$ for some of the lowest harmonics in the lower panel of Fig.~\ref{fig:plot3}.

We use the expected LISA spectral sensitivity $S(f)$~\cite{amaroseoane2017laser,Robson:2018ifk} and include noise arising from the galactic and extragalactic foregrounds~\cite{Lewicki:2021kmu}. We considered sources at $z_s=0.5$ and at $z_s=1$ with $f_2 = 2\,$mHz. To assess the detectability of the shifted timbre effect, we consider only the harmonics $n = 2-7$ that dominate the signal for $e \lesssim 0.6$, as seen in the lower panel of Fig.~\ref{fig:plot3}. We determine how much ${\rm SNR_2}$ and $e$ are required to detect the shifted timbre. We use the threshold $\Delta\chi^2 > 8.02$ corresponding to $2\sigma$ CL (Confidence Level) for a template with three parameters. The results are shown in the upper panel of Fig.~\ref{fig:plot3}. For example, for $z_s=0.5$ and $e=0.3$ the dark timbre is detected if ${\rm SNR_2}\gtrsim 3000$ while for $e=0.5$ it is enough to have ${\rm SNR_2}\lesssim 500$. For $z_s=1$, approximately factor four smaller ${\rm SNR_2}$ is sufficient. An example of a source with approximately these properties could be an IMBH binary with $\mathcal{M}=\mathcal{O}(10^3/M_{\odot})$ and $\eta\sim0.1$. In scenarios with light supermassive BH seeds, the number of detectable binaries of this type would be $\mathcal{O}(100)/{\rm year}$ and they would enter the LISA band at $\mathcal{O}({\rm years})$ before merger~\cite{Ellis:2023iyb}. For some time, they would be approximately monochromatic with SNR of $\mathcal{O}(10^3)$. As shown in~\cite{Porter:2010mb}, also the referred eccentricities are realistic. 

The precision at which the amplitudes of a harmonic $i$ can be resolved is approximately
\be
    \frac{\sigma_{A_i}}{A_i}\sim\frac{1}{{\rm SNR}_i}\, ,
\ee
which is easy to show from Fisher analysis~\cite{Poisson:1995ef}. This implies that a higher ${\rm SNR}_2$ is necessary to resolve the relative amplification factors $F_{i,j}$ than to detect the dark timbre. Optimistically, if a population of eccentric binaries is detected with ${\rm SNR}_i \sim 10^3$ it would be possible to reconstruct the PDFs (Probability Density Functions) of $F_{i,j}$ and probe the HMF down to $M_{\rm v} \ll 10\,M_\odot$. However, already the detection of one sufficiently eccentric binary with ${\rm SNR}_2 \gtrsim 10^2$ at $z_s=0.5$ would tell about the existence of halos with $M_{\rm v} \lesssim 10\,M_\odot$. 

We note that since the dark timbre effect is very small, corrections beyond the Newtonian order~\cite{Peters:1963ux} in the different harmonics coefficients may become relevant, although we assume that the binaries are very far from merging, furthermore considering more realistic waveforms with higher modes would potentially help in the detectability~\cite{Caliskan:2022hbu}. Moreover, detecting the dark timbre requires identifying the frequency harmonics. LISA is expected to observe several nearly monochromatic GWs simultaneously at different frequencies. Some of these GWs may be harmonics from an eccentric IMBH binary, while others may originate, e.g., from galactic WD binaries~\cite{Korol:2017qcx}. Consequently, resolving which of these GWs are harmonics emitted by the same source can be challenging. The effects of higher modes and the identification problem of the frequency harmonics are left for future studies.

\section{Conclusions} 

We have estimated for the first time the integrated lensing effect of light halos on the mid-frequency (mHz) GW signals. We have proposed to test these effects using the different harmonics of an eccentric binary. We have shown that the wave optics effects due to the low-mass DM halos, $M_{\rm v} \lesssim 10\,M_{\odot}$, induce frequency-dependent changes in the amplitude and phase of the harmonics -- the timbre of the signal. This shifted timbre is detectable in the 7 dominant harmonics of the signal for signal-to-noise ratios between $100 - 10^3$  if the binary eccentricity is $e = 0.3 - 0.6$ and it is $z_s=0.5$. If such binaries exist, LISA could probe the dark timbre. This would open a new avenue to test low-mass DM halos and thus provide new insights into the nature of DM. 

\begin{acknowledgments}
\vspace{5pt}\noindent\emph{Acknowledgments --} We thank Luca Marzola and Hardi Veerm\"ae for useful suggestions. This work was supported by the Estonian Research Council grants PRG803, PSG869, RVTT3 and RVTT7 and the Center of Excellence program TK202. The work of V.V. was partially supported by the European Union's Horizon Europe research and innovation program under the Marie Sk\l{}odowska-Curie grant agreement No. 101065736.
\end{acknowledgments}

\bibliography{refs}

\appendix

\section{Validity of the Born and eikonal approximation}
\label{appA}

In the following, we make explicit the assumptions behind the formula of the amplification factor $F$ used in the main text and showcase their validity.

To compute the scatter of a GW from the DM halo we will consider the metric around the halo to be given by 
\be
    \td s^2 = -(1+2U(\vect{x})) \td t^2 + a^2 (1-2U(\vect{x})) \td \vect{x}^2 \,,
\ee
where $U$ is the gravitational potential of the halo and $a$ denotes the scale factor that we consider as a constant fixed by the halo redshift. Under the Born approximation, we consider only the leading order terms in $U\ll 1$. Moreover, we neglect the change in the polarization of the GW which is of the order $U\ll 1$ and consider only its amplitude $\phi(t,\vect{x})$. Then, the Fourier transform $\Tilde{\phi}(\omega,r)$ of the scalar wave follows the wave equation
\be\label{eq:mov}
    \left(a^{-2} \nabla^2 + \omega^2\right) \Tilde{\phi}(\omega,\vect{x}) = 4\omega^2\,U\Tilde{\phi}(\omega,\vect{x}) \,. 
\ee
We want to express the solution as 
\be \label{eq:sol}
    \Tilde{\phi}(\omega,\vect{x}) = F(\omega,\vect{x})\Tilde{\phi}_{\rm NL}(\omega,\vect{x}) \,,
\ee
where the $\tilde{\phi}_{\rm NL}(\omega,\vect{x})$ is the solution of Eq.~\eqref{eq:mov} for $U = 0$. Plugging~\eqref{eq:sol} into the equation of motion~\eqref{eq:mov} we get that the amplification function $F(\omega,r)$ has to fulfill 
\be \label{eq:F1}
    \frac{1}{(a\, r)^2}\nabla^2_{\theta} F+\frac{1}{a^2}\frac{\partial^2 F}{\partial\, r^2}+\frac{2i\omega}{a}\frac{\partial F}{\partial r}=4\omega^2 U F \,.
\ee
Assuming that 
\be \label{eq:eiko}
    \left|\frac{1}{a^2}\frac{\partial^2 F}{\partial\, r^2}\right|\ll \left|\frac{2i\omega}{a}\frac{\partial F}{\partial r}\right| \,,
\ee
Eq.~\eqref{eq:F1} simplifies to
\be\label{eq:schr}
    \frac{i}{a}\frac{\partial F}{\partial r}=\frac{1}{2\omega (a\, r)^2}\nabla^2_{\theta} F+2\omega U F \,.
\ee
which is a Schrödinger equation that can be solved with the path integral approach that gives the formula of $F$ we use throughout the paper.

\begin{figure}
    \centering
    \includegraphics[width=0.96\columnwidth]{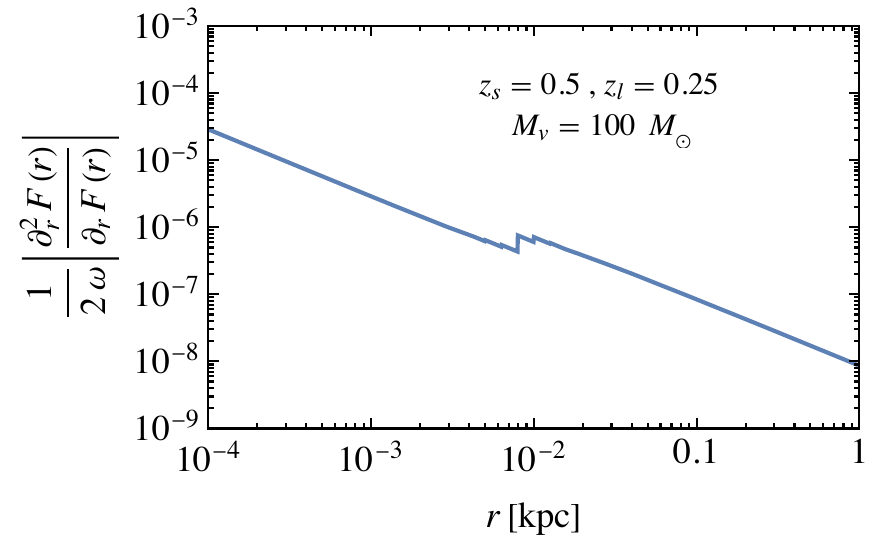}
    \caption{\textit{Upper panel:} Comparison of the second and first derivatives of $F$. \textit{Lower panel:} Values of $U$ for the typical impact parameters that contribute to the dark timbre.}
    \label{fig:apen_1}
\end{figure}

In the lower panel of Fig.~\ref{fig:apen_1} we plot the NFW gravitational potential,
\be
    U(r) = -\frac{4\pi\rho_0r_s^3}{r}\log{\left(1+\frac{r}{r_s}\right)} \,,
\ee
to show that the Born approximation is valid for the impact parameters we typically consider for the dark timbre. In the upper panel we demonstrate the validity of the eikonal approximation.

\section{Multilens computation}
\label{appB}

\begin{figure}
    \centering
    \includegraphics[width=0.98\columnwidth]{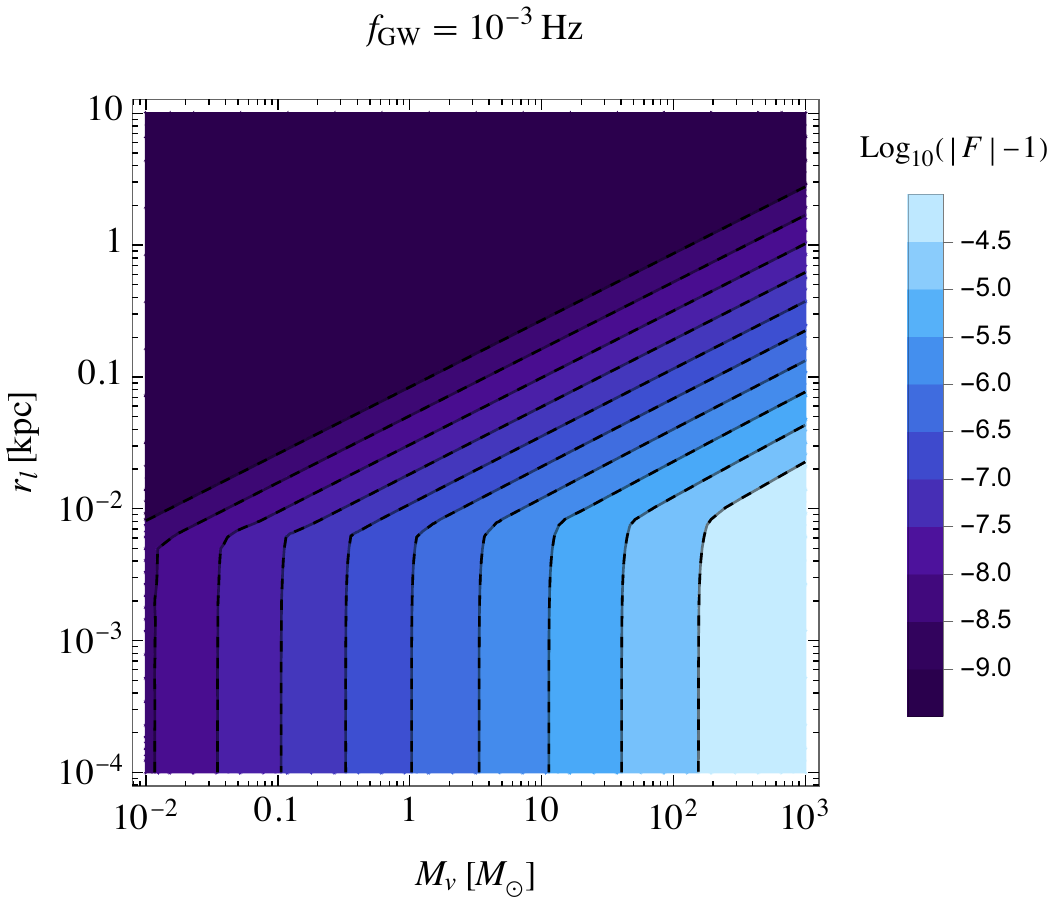}
    \caption{Comparison between the amplification function for a single lens computed without (solid contours) and with (dashed contours) the expansion in $\omega \psi_j$.}
    \label{fig:apen_2}
\end{figure}

The solution of Eq.~\eqref{eq:schr} is given by the path integral
\be
    F = \int {\mathcal D}\vect{\theta} \, e^{i\int_0^{r_{N+1}}\td r \,\mathcal{L} } \,, 
\ee
where $\vect{\theta}=\theta(\cos{(\phi)},\sin{(\phi)})$ parametrizes the paths that the GW can take from the source at $r=0$, $\vect{\theta} = 0$ to the observer at $r=r_{N+1}$, $\vect\theta = \vect\theta_{N+1} = 0$ and
\be
    \mathcal{L} = \frac12\omega (a r)^2 |\dot{\vect\theta}(r)|^2 - 2\omega U \,.
\ee
We divide the path into $N$ slices perpendicular to the radial vector from the source to the detector at distances $r= r_j$ of thickness $\epsilon$, so that $r_{N+1} = (N+1)\epsilon$. Some of these slices include a lens. In the thin lens approximation, we project the gravitational potentials of each lens to a plane,
\be
    U(r,\vect\theta) = \frac12 \sum_{j=1}^N \psi_j(\vect\theta) \delta(r-r_j)
\ee
and consider paths that change direction only at the lens planes,
\be
    |\dot{\vect\theta}(r)| = \sum_{j=1}^N \Theta(r-r_j) \Theta(r_{j+1}-r) \frac{r_j r_{j+1}}{a r \epsilon} |\vect\theta_{j+1}-\vect\theta_j| \,,
\ee
where $\vect\theta_j$ denotes the position where the path crosses the $j$th lens plane and $r_{i,j} \equiv r_ir_j/(r_i-r_j)$. Then the integrated Lagrangian can be expressed as 
\be
    \int_0^{r_{N+1}}\td r\, \mathcal{L} = \omega \sum_{j=1}^{N} \left[ \epsilon \frac{r_j r_{j+1}}{2} \left|\frac{\vect\theta_{j}-\vect\theta_{j+1}}{\epsilon}\right|^2 - \Psi_j(\vect\theta_j)\right] \,.
\ee
and amplification function as
\bea \label{eq.F_comp}
    F = \!\int &\!\prod_{j=1}^N \frac{\omega r_j r_{j+1}}{2\pi i \epsilon} \td^2 \vect\theta_j \\
    &\times \exp\!\bigg[ i\omega \bigg( \frac{r_j r_{j+1}}{2 \epsilon} |\vect\theta_{j+1} - \vect\theta_{j}|^2 - \psi_j(\vect\theta_j)\bigg)\bigg] .
\eea
Here $\psi_j = 0$ for those slices that don't include a lens. Assuming that for all relevant paths $\omega \psi_j \ll 1$, we can expand to the lowest order in powers of $\omega \psi_j$:
\bea \label{eq.F_comp}
    F \approx & \int \!\prod_{j=1}^N \frac{\omega r_j r_{j+1}}{2\pi i \epsilon} \td^2 \vect\theta_j \exp\!\bigg[ i\omega \frac{r_j r_{j+1}}{2 \epsilon} |\vect\theta_{j+1} - \vect\theta_{j}|^2\bigg] \\
    &\qquad \times \left[ 1 - i \omega \psi_j(\vect\theta_j) \right]  \\
    \approx & 1 - i \omega \sum_{l=1}^N \int \frac{\omega r_l r_{l+1}}{2\pi i \epsilon} \td^2 \vect\theta_l \left[\prod_{j\neq l} \frac{\omega r_j r_{j+1}}{2\pi i \epsilon} \td^2 \vect\theta_j\right] \\
    &\qquad \times \psi_l(\vect\theta_l) \exp\!\bigg[ i\omega \sum_{j=1}^N \frac{r_j r_{j+1}}{2 \epsilon} |\vect\theta_{j+1} - \vect\theta_j|^2 \bigg] .
\eea
The integrals for $j<l$ give unity and~\cite{Nakamura:1999uwi}
\bea
    &\sum_{j=l}^N \frac{r_j r_{j+1}}{2 \epsilon} |\vect\theta_{j+1} - \vect\theta_j|^2 = \epsilon \frac{r_l r_{N+1}}{r_{l,N+1}} |\vect\theta_{N+1} - \vect\theta_l|^2 \\
    &\qquad\qquad\qquad + \sum_{j=l+1}^N r_j^2 \frac{r_l r_{j+1}}{r_{l,j}} |\vect\theta_j - \vect u_{l,j}|^2 \,,
\eea
where $r_{l,j} = r_l - r_j$ and $\vect u_{l,j} = [r_l \vect\theta_j + (j-l) r_{j+1} \vect\theta_{j+1}]/(j r_{l,j+1})$. So, all integrals except that over $\vect\theta_l$ are trivial and give $\epsilon r_{N+1}/(r_{l+1} r_{l,N+1})$. This results in
\bea
    F &\approx 1 - i \omega \sum_{l=1}^N \int \frac{\omega r_l r_{N+1}}{2\pi i r_{l,N+1}} \td^2 \vect\theta_l \psi_l(\vect\theta_l) \\ &
    \qquad \qquad \qquad \times \exp\!\bigg[ i\omega \frac{r_l r_{N+1}}{r_{l,N+1}} |\vect\theta_{N+1} - \vect\theta_l|^2 \bigg] \\
    &\approx 1 - \sum_{l=1}^{N} (F_j-1) \,.
\eea
In Fig.~\ref{fig:apen_2}, we show that the expansion in $\omega \psi_j \ll 1$ works well for the NFW lenses. The solid and dashed contours are computed without and with the expansion and their excellent agreement is evident.

\end{document}